\documentclass{jfm}

\usepackage{amsmath}
\usepackage{color}
\usepackage[normalem]{ulem}
\usepackage{cancel}
\usepackage{esint}
\usepackage{hyperref}
\usepackage{graphics}
\usepackage{natbib}
\DeclareMathOperator*{\res}{Res}
\DeclareMathOperator{\real}{Re}
\DeclareMathOperator{\imag}{Im}

\DeclareMathOperator{\acosh}{acosh}

\begin{document}

\newtheorem{lemma}{Lemma}
\newtheorem{corollary}{Corollary}

\newcommand\pb{{^+\!b}}
\newcommand\pc{{^+\!\!c}}
\newcommand\PW{90mm}

\shorttitle{Singularities, Invariants and Virtual Fluid} 
\shortauthor{A. I. Dyachenko et al} 

\title{Free Surface in 2D Potential Flow: Singularities, Invariants and Virtual Fluid}

\author
 {
 A. I. Dyachenko\aff{1,2,3}, 
 S. A. Dyachenko\aff{4,\corresp{\email{sergeydy@buffalo.edu}}}
 \and
  V. E.  Zakharov\aff{1,2,5}
}
\affiliation
{
\aff{1}
Landau Institute for Theoretical Physics, 142432, Chernogolovka, Russia
\aff{2}
Skolkovo Institute of Science and Technology, Moscow, Russia
\aff{3}
National Research University Higher School of Economics,
101000, Myasnitskaya 20, Moscow, Russia
\aff{4}
Department of Mathematics, University at Buffalo, SUNY Buffalo, NY, 14260, USA
\aff{5}
Department of Mathematics, University of Arizona, Tucson, AZ, 85721, USA
}
\maketitle

\begin{abstract}
We study a 2D potential flow of an ideal fluid with a free surface with decaying conditions 
at infinity. By using the conformal variables approach, we study a particular solution of Euler
equations having a pair of square--root branch points in the conformal plane, and find that 
the analytic continuation of the fluid complex potential and conformal map define a flow in the 
entire complex plane, excluding a vertical cut between the branch points. The expanded domain is 
called the ``virtual'' fluid, and it contains a vortex sheet whose dynamics is equivalent to 
the equations of motion posed at the free surface. The equations of fluid motion are analytically 
continued to both sides of the vertical branch cut (the vortex sheet), and additional time--invariants 
associated with the topology of conformal plane and  Kelvin's theorem for virtual fluid are explored. 
We called them ``winding'' and virtual circulation. This result can be generalized to a system of many 
cuts connecting many branch points, and resulting in a pair of invariants for each pair of branch 
points. We develop an asymptotic theory that shows how a solution originating from a single vertical 
cut forms a singularity at the free surface in infinite time, the rate of singularity approach is 
double-exponential, and supercedes the previous result of the short branch cut theory with 
finite time singularity formation.

\indent The present work offers a new look at fluid dynamics with free surface by unifying the problem of 
motion of vortex sheets, and the problem of 2D water waves. A particularly interesting question that 
arises in this context is whether instabilities of the virtual vortex sheet are
related to breaking of steep ocean waves when gravity effects are included.
\end{abstract}


\section{Introduction}

Motion of ideal fluid with a free surface is one of the oldest problems in applied mathematics, and 
emergence of complex analysis can be attributed to the study of potential flows in 2D. A fluid 
flow that is coupled to the motion of a free boundary as in the motion of waves at the surface 
of ocean becomes particularly rich and complex. Many classical problems in nonlinear science are
tied to the dynamics of ocean surface: the nonlinear Schr\"oedinger equation (NLSE) and the 
Korteweg--de--Vries equation (KdV) both can be derived as an approximation to water wave motion under 
distinct assumptions. Yet both models share a particularly striking property: integrability. 
Integrable systems are quite rare, and one of their special features is dynamics that is uniquely 
determined by a set of integrals of motion and phases, also referred to as action and angle variables
see e.g.~\citep{kolmogorov1954conservation}. The state of an integrable system at a given time can be 
determined by means of the inverse scattering technique, see the works on integrable systems~\citep{gardner1967method,zakharov1971korteweg,shabat1972exact,zakharov1974scheme,ablowitz1974inverse,zakharov1979integration}. 

At present many nonlinear systems have been discovered, yet integrability of the full water wave system
remains elusive. The search for integrability in water waves is a long standing problem, and it was
proven that if it indeed exists it is of a special kind. The Ref.~\citep{dyachenko1995five} applies 
Zakharov--Schulman technique to water waves and shows that the fluid dynamics is not integrable with 
time--invariant spectrum. Nevertheless, 
new nontrivial integrals of motion have been discovered~\citep{Tanveer1993,dyachenko2019poles,lushnikov2021poles,Dyachenko2021short}) 
that suggest presence of hidden structure and suggesting integrability in a broader sense. 
The new integrals of motion are related to contour integrals in the analytic continuation of the fluid domain,
the ``virtual fluid'' (also sometimes referred to as phantom and/or unphysical), which is an abstraction defining a fluid flow in a maximally extended domain where 
the analytic functions defining the flow reach their natural boundaries of analyticity. 
In the preceding work~\citep{dyachenko2019poles} the authors found that if a singularity of complex velocity in
the virtual fluid is a pole, then its residue is a time invariant. Nevertheless, appearance of isolated 
singularities in a generic flow is observed under very special circumstances~\citep{PK45,G45,ZD96PD}. 
Even then isolated singularities alone are incompatible with a fluid flow with free surface, 
see the work~\citep{lushnikov2021poles}. Exact solutions originally found by Dirichlet and described in
the reference~\citep{LonguettH1972} are second order curves and also contain square--root branch points. 
A notable exception is a classical work of~\citep{Crapper1957} who discovered the traveling wave on a free surface subject to forces of surface tension; the Crapper waves 
are one of the few exact solutions and their singularities are isolated poles, the work~\cite{crowdy2000new} discusses the mathematical framework to construction of flows with surface tension and rational solutions in particular. The recent work by~\citep{dyachenko2019stokes} and the following theoretical proof by~\citep{hur2020exact} show that 
Crapper waves also occur in the vanishing gravity limit for waves over a shear current.

The importance of square-root branch points for potential fluid flow has been first discovered in the work~\citep{Tanveer1993}.
In the work~\citep{baker2011singularities} the authors concluded that 
square--root branch point approaches the fluid region when 
a breaking gravity wave becomes overhanging. Formation
of square--root singularity and whitecapping was also conjectured in~\citep{DyachenkoNewell2016}.
The work~\citep{castro2013finite} is a study of the 
formation of multivalued surface through the ``splash'' mechanism: a scenario in which a 
free surface becomes self-intersecting; the authors show that splash singularity 
may appear in finite time while originating from perfectly smooth initial datum. 
The origin of appearance of branch points is the complex Hopf 
equation~\citep{kuznetsov1993surface,karabut2014unsteady,karabut2020application} that governs fluid motion in some approximation. 
Many nontrivial flows are described by square--root branch points see for example the
works~\citep{ZakharovEtAl1996,DyachenkoEtAl1996,zakharov2020integration,liu2021search} for the 
study of free surface with decaying 
boundary conditions, and the recent work~\citep{Dyachenko2021short} for periodic case. To further 
the case, a periodic wave that is traveling on a free surface also known as the Stokes wave has 
square-root branch points as found in the work~\citep{grant1973singularity} and numerical study of its 
singularities in~\citep{dyachenko2014complex, dyachenko2016branch} by means of rational approximation~\citep{alpert2000rapid}, 
and the singularity of the limiting Stokes wave is conjectured to be the result of coalescence of 
multiple square--root branch points, see the work~\citep{lushnikov2016structure}.

2D fluid flows can be studied using the conformal variables approach, which was 
first introduced in the $19$th century. The pioneering work of~\citep{Stokes1880} 
on traveling periodic waves discusses conformal mapping in the context of $120^{\circ}$ at the crest, see 
also a recent review paper~\citep{haziot2021traveling} and references therein. The problem of finding 
standing waves is more mysterious, due to its complicated temporal dynamics. A recent
work~\citep{wilkening2021traveling} discusses a technique for construction of traveling--standing
waves that bridge the gap between traveling and standing waves. 
The first application of conformal mapping to time-dependent flows can be traced to the 
work of~\citep{ovsyannikov1973dynamika}, 
that followed a result of~\citep{Zakharov1968} who discovered the canonical Hamiltonian variables  
for fluid flow consist of the free surface and the velocity potential on it. 

The conformal mapping technique is not always the most convenient way to study water waves 
numerically, and we will refer the reader to the work~\citep{wilkening2015comparison} 
for other highly efficient methods for 2D fluid dynamics. The recent works~\citep{arsenio2020vortex,ambrose2021numerical} discuss novel methods for simulating Euler 
equation in 2D, which generalize to 3D water waves. 
The conformal mapping technique is discussed in the works~\citep{tanveer1991singularities,Tanveer1993}, ~\citep{ZakharovEtAl1996},~\citep{Dyachenko2001} and successfully applied numerically by many authors, see e.g. 
the works~\citep{ZakharovEtAl2002,DyachenkoNewell2016}.

In the present work we develop an exact theory of a potential 2D flow in Euler equations with 
a free surface. The theory describes a particular solution that carries a pair of square--root 
branch points in the analytic continuation of the complex velocity and the conformal map, and 
offers a pair of newly discovered integrals of motion, the ``windng'' and ``circulation'' of virtual
fluid. Asymptotic theory is developed that shows double--exponential approach of a square--root 
branch point to the fluid domain that suggests formation of singularity in infinite time, which is 
distinct from the short branch--cut theory developed in preceding works.

\section{The Square-Root Branch Cut}
\begin{figure}
\includegraphics[width=0.95\textwidth]{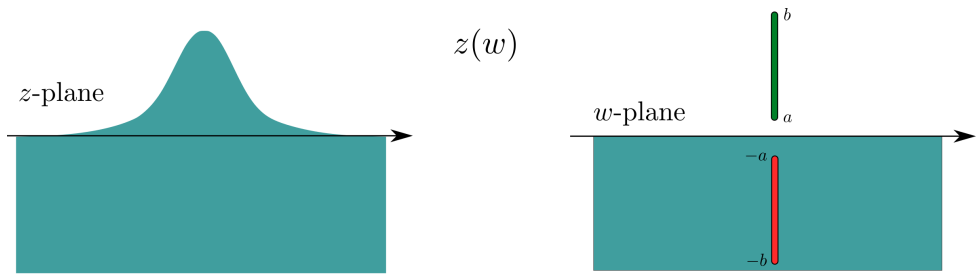}
	\caption{(Schematics of conformal map $z$) The fluid region (green) is mapped to the lower complex plane $\imag w < 0$,
	the branch cut of $R$, $V$ is marked in blue. The branch cuts for complex conjugate functions are marked red.}
	\label{fig:ZMap}
\end{figure}
The nonlinear equations for dynamics of free surface of 2D fluid, written in conformal variables, are known since~\citep{ovsyannikov1973dynamika,tanveer1991singularities}. We consider the form of these equations given in the work~\citep{Dyachenko2001}:
\begin{eqnarray}\label{RV}
&&\dot R = i\left(UR' - U'R\right)\cr
&&\dot V = i\left(UV' - B'R\right)
\end{eqnarray}
The functions $R(w,t)$, $V(w,t)$, $U(w,t)$, and $B(w,t)$ are analytic with respect to complex variable $w=u+iv\in\mathbb{C}^-$ (the lower half-plane).
Here $U(w,t)$ and $B(w,t)$ are analytically continued from the real line, where they are given by the relations:
\begin{equation}
U(u,t) = \hat P^-[\bar V(u,t) R(u,t)+V(u,t)\bar R(u,t)],\hspace{20pt} B(u,t) = \hat P^-[V(u,t) \bar V(u,t)].
\end{equation}
Here $\hat P^-$ is the projection operator from the real axis to
the lower half-plane. Given on the real axis
\begin{equation}\nonumber
\hat P^- =\frac{1}{2}(1+i\hat H), \hspace{10pt}\hat H \hspace{6pt}
\mbox{- is the Hilbert tranform:}\hspace{6pt} 
\hat Hf = p.v.\frac{1}{\pi}\int_{-\infty}^{\infty}
\frac{f(u',t)du'}{u'-u}.
\end{equation}
Let 
\begin{align}
R(w,t) = 1 + \rho(w,t),\quad\mbox{where}\quad \rho(w\to \infty) \to 0
\end{align}
The four analytic functions have square--root branch points at $w = ia(t)$ and $w = ib(t)$, and thus can be expressed by means of the Cauchy integral formula with the clockwise contour orientation:
\begin{align}
    \rho(w,t) = \frac{1}{2\pi i} \oint\limits_{[ia,ib]} \frac{\rho(s,t)\,ds}{s - w} = 
    \frac{i}{2\pi} \lim\limits_{\varepsilon \to 0^+}\int\limits_{ia}^{ib} \frac{\left[\rho(s+\varepsilon) - \rho(s - \varepsilon)\right]ds}{s-w}, \label{CtypeR}
\end{align}
and we denote $\rho(s\pm\varepsilon) \to \rho^{\pm}(s)$ as $\varepsilon \to 0^+$. Let $\tau = -is\in[a,b]$, then we rewrite the formula as follows:
\begin{align}
    \rho(w,t) = \frac{1}{2\pi}\int\limits_a^b \frac{\left[\rho^+(i\tau) - \rho^-(i\tau)\right]d\tau}{-i\tau+w}
\end{align}
and define jumps on the cut for $\rho$ and $V$:
\begin{align}
    r(\tau) = \frac{\rho^+(i\tau) - \rho^-(i\tau)}{2i} \quad\mbox{and}\quad
    v(\tau) = \frac{V^+(i\tau) - V^-(i\tau)}{2i}
\end{align}
and note that $r(a) = r(b) = 0$ and $v(a) = v(b) = 0$. The associated analytic functions are then given by:
\begin{align}
    R(w,t) = 1 - \frac{1}{\pi} \int\limits_a^b \frac{r(\tau)\,d\tau}{\tau + iw}\quad\mbox{and}\quad
    V(w,t) = -\frac{1}{\pi} \int\limits_a^b \frac{v(\tau)\,d\tau}{\tau+iw}
\end{align}
We also introduce the two additional functions from the following relations to introduce $U$ and $B$:
\begin{align}
    \bar R V + R \bar V &= V + \bar V + \frac{1}{\pi^2}\iint \frac{\left[\bar r(\tau')v(\tau) + r(\tau)\bar v(\tau')\right]d\tau d\tau'}{(\tau + iw)(\tau' - iw)}, \\
    \bar V V &= \frac{1}{\pi^2}\iint\frac{v(\tau)\bar v(\tau') d\tau d\tau'}{(\tau + iw)(\tau' - iw)}
\end{align}
We refer the reader to the derivation in Appendix~\ref{AppendixA:U_and_B} for derivation of the following 
formulas:
\begin{align}
    U(w) = -\frac{1}{\pi}\int\frac{u(\tau)d\tau}{\tau + iw}\quad\mbox{and}\quad
    B(w) = -\frac{1}{\pi}\int\frac{b(\tau)d\tau}{\tau + iw}\label{UB}
\end{align}
where $u(\tau) = v(\tau)\bar R(i\tau) + r(\tau) \bar V (i\tau)$ and $b(\tau) = v(\tau)\bar V(i\tau)$ for $\tau\in[a,b]$.

\subsection{Boundary values of analytic function at the cut}
Before proceeding, one must be able to evaluate the complex analytic function by its associated jump on the cut. Given $\varepsilon > 0$, one finds that 
\begin{align}
    R(i\tau \pm \varepsilon) = 
    1 - \frac{1}{\pi}\int\limits_a^b
    \frac{r(\tau')d\tau'}{\tau' - \tau \pm i\varepsilon}\quad\mbox{and}\quad
    V(i\tau \pm \varepsilon) = 
    - \frac{1}{\pi}\int\limits_a^b
    \frac{v(\tau')d\tau'}{\tau' - \tau \pm i\varepsilon}
\end{align}
and may apply the Sokhotskii--Plemelj theorem to obtain:
\begin{align}
    R^{\pm}(i\tau) &= \lim\limits_{\varepsilon\to 0^+} R(i\tau \pm \varepsilon) = 1 - \left(\hat H \mp i\right) r(\tau), \\  V^{\pm}(i\tau) &= \lim\limits_{\varepsilon\to 0^+} V(i\tau \pm \varepsilon) = -\left(\hat H \mp i\right) v(\tau),
\end{align}
where we have defined the integral operator $\hat H$ as follows:
\begin{align}
    \hat H f(\tau) = \frac{1}{\pi}v.p.\int\limits_a^b\frac{f(\tau')d\tau'}{\tau'-\tau}
\end{align}
Similar relations hold for the functions $U$ and $B$.

\subsection{Equations of motion on the cut}

The equations of motion in $w$--plane have been derived previously, and are given by:
\begin{align}
    \partial_t R &= i\left(U R_w - U_w R\right), \\
    \partial_t V &= i\left(U V_w - B_w R\right)
\end{align}
and can be written in terms of the jumps of the associated functions as follows:
\begin{eqnarray}\label{rv}
    r_t +  u\hat H r' + r'\hat H u + u'(1-\hat Hr) - r\hat Hu' &=& 0 \cr
    v_t + u \hat H v' + v'\hat H u + b'(1-\hat Hr) - r\hat Hb' &=& 0
\end{eqnarray}

%
%

Given a square-root branch point in $z(w) = \sqrt{w-ia}$, the function $R = 1/z_u \sim \sqrt{w-ia}$ vanishes at the branch points. Similarly, it is trivial to show that $R(w,t)$ has zeros at the branch points $ia$ and $ib$:
\begin{align}
    R(ia) = R(ib) = 0\quad\mbox{and}\quad R(w\to -i\infty) \to 1 \label{nine} 
\end{align}
as the solution evolves the branch points do not vanish, but move in the complex plane see~\citep{Dyachenko2021short}.

One can use the shifted Chebyshev basis to efficiently represent the complex analytic functions with a 
pair of square--root singularities. We introduce the center of the cut, $c(t) = \frac{a(t)+b(t)}{2}$, and 
its half-length, $l(t)= \frac{b(t)-a(t)}{2}$, and use $f_n(w,t)$ as a basis for expansion of the analytic 
functions $R(w,t)$ and $V(w,t)$. It is convenient to work with the variable $\xi(w,t) = \frac{c+iw}{l}$, then:
\begin{align}
    f_n(\xi) &= \left[\xi - \sqrt{\xi^2  - 1}\right]^n = (-1)^n\left[T_n(-\xi) + U_{n-1}\left(-\xi\right)\sqrt{\xi^2 - 1}\right] , \label{analytic_Cheb}
\end{align}
where $T_n(\xi)$ and $U_{n-1}(\xi)$ for $n = 1,2,\ldots$ are the Chebyshev polynomials of the first and 
the second kind respectively. Note that $w\in[ia,ib]$ is mapped to $\xi\in[-1,1]$ for convenience.

\subsection{Expansion of $R(w,t)$ and $V(w,t)$}
\begin{figure}
\includegraphics[width=0.95\textwidth]{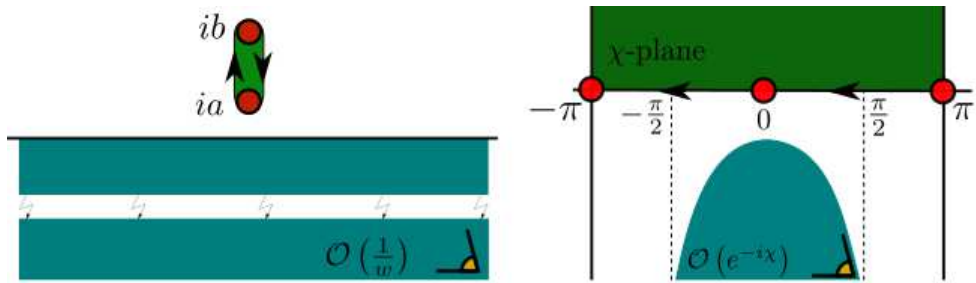}
	\caption{(Schematics of $\chi$-map) A periodic strip $-\pi \leq \real \chi< \pi$ and 
	$\imag \chi < 0$ (right) is  
	mapped to $w\in\mathbb{C}\cap[ia,ib]$ (left), and the $\imag \chi > 0$ (right) is mapped to the second 
	sheet of $w$-plane. (Left Panel) The blue region is mapped to the fluid domain (the real fluid), and the upper halfplane marked white excluding the cut is the virtual fluid in the first sheet outside the cut. The red circles mark the location of the branch points, and arrows indicate the positive orientation for a Cauchy-type integral~\eqref{CtypeR}. An observer marked with an eye that is located far away from the cut (large distance from real line is illustrated by the lightning symbols) sees the dominant term of a far field $1/w$ whose coefficient is a motion integral defined in~\eqref{farfield}.
	The green domain observed on the left is the virtual fluid of 
	the second Riemann sheet which is seen through the cut.
	(Right Panel) After the conformal map~\eqref{chi1} the image of a periodic contour around the cut is mapped to the interval $[-\pi,\pi]$, and the Chebyshev function basis~\eqref{analytic_Cheb} becomes the standard Fourier basis~\eqref{analytic_Fourier}. The image of the real fluid is located in the lower complex plane enclosed within vertical asymptotes $\chi = \pm \frac{\pi}{2}$ (dashed lines), and the virtual fluid from the first sheet is mapped to the white region in the lower halfplane. The second Riemann sheet is unfolded into the upper halfplane of the $\chi$-plane. The singularities in the upper half-plane of $\chi$ are located in the second Riemann sheet of $w$-plane, and can be studied numerically and theoretically.} 
	\label{fig:ChiMap}
\end{figure}

The complex velocity, $V$, and $R$ may be expanded in the form:
\begin{align}
    V(\xi,t) = \sum\limits_{k=1}^{\infty} v_k(t) f_k(\xi)\quad\mbox{and}\quad
    R(\xi,t) = 1 + \sum\limits_{k=1}^{\infty} r_k(t) f_k(\xi). \label{cheb_exp}
\end{align}
In order to satisfy the conditions~\eqref{nine} we have two additional constraints on the coefficients of $R$ that
must be satisfied for any $t>0$:
\begin{align}\nonumber
    \sum\limits_{k=1}^\infty r_k(t) = \sum\limits_{k=1}^{\infty} (-1)^k r_k(t) = -1
\end{align}
It is convenient to rewrite the series defined in relations~\eqref{cheb_exp} and~\eqref{analytic_Cheb}
using the following substitution
\begin{align}
    \xi = \cos{\chi}\quad\mbox{or, equivalently}\quad w = ic - il\cos{\chi}, \quad \chi = \eta +i \zeta \label{chi1}
\end{align}
where $\chi\in[-\pi,\pi]$ is mapped to $\xi \in [-1,1]$. Moreover, the relation~\eqref{chi1} defines
a conformal mapping for complex $w$, such that $\chi(w\in \mathbb{C})\to \mathbb{C^-}$ (see also
illustration in Fig~\ref{fig:ChiMap}), and the free surface (blue-white boundary) is located at: 
\begin{equation}\label{sqrt}
    \chi(u) = i\ln\left[ \xi - \sqrt{\xi^2 - 1}\right], \,\,\,\mbox{where}\,\,\,\xi = \frac{c+iu}{l}.
\end{equation}
Note: square root in (\ref{sqrt}) is so that 
$\sqrt{a^2 e^{i\phi}}=|a|e^{i\frac{\phi}{2}}$.

The components of the Chebyshev series:
\begin{align}
    f_n(\xi) &= \left[\xi - \sqrt{\xi^2  - 1}\right]^n = e^{-in\chi} , \label{analytic_Fourier}
\end{align}
becomes the standard Fourier series in $\chi$ variable in negative Fourier harmonics. Note that 
all functions $R$, $V$, $U$, $B$ {\bf and their complex conjugates in the $w$-plane} can be expanded in 
the series~\eqref{analytic_Fourier} with $n \geq 0$. 
\begin{align}\label{fourier_exp}
    V(\chi,t) = \sum\limits_{k=1}^{\infty} v_k(t) e^{-ik\chi}\quad\mbox{and}\quad
    R(\chi,t) = 1 + \sum\limits_{k=1}^{\infty} r_k(t) e^{-ik\chi}.
\end{align}

In order to determine the functions $U$ and $B$ we will first evaluate evaluate the auxiliary functions $u(\tau)$ and $b(\tau)$ at the cut. 
Then we will use the Sokhotskii-Plemelj Thm to find the values of $U$ and $B$ at 
the cut, $\chi = \eta + i0 \in[-\pi,\pi]$.
The values of $R$ and $V$ at the reflection of the contour around the cut in the lower halfplane of $w$ can be determined in the $\chi$-plane. The reflection of the cut is located on the negative imaginary axis at $\chi = 0 + i\zeta_{cc}(\tau)$ given by the relation:
\begin{align*}
    &\cos i\zeta_{cc} = \frac{c + \tau}{l},  \\
    &\cos i\zeta_{cc} = 2 + 2\frac{a}{l} - \xi =  2 + 2\frac{a}{l} - \cos{\eta} 
\end{align*}
where $\tau \in [a,b]$ (see also the formula~\eqref{UB}). Note that the righthand side
of the formula is strictly greater than $1$, and the values of $\zeta_{cc}$ are purely
real and negative, given our choice of the mapping $\chi(w)$. Solving for $\zeta_{cc}$,
we find:
\begin{align*}
    \zeta_{cc}(\xi) = -\ln \left[ y + \sqrt{y^2 - 1} \right] = -\acosh{y}, 
\end{align*}
where $y = 2 + 2\frac{a}{l} - \xi$. The functions $\bar R$ and $\bar V$ can then be
found on the reflection of the cut $w\in[-ia,-ib]$ (or equivalently $\chi\in i\zeta_{cc}$) by evaluating the formulas:
\begin{align*}
    &\bar R = 1 - \pi \sum\limits_{k=1} k_n \left(\sqrt{y^2 -1} -y\right)^k = 1 - \pi \sum\limits_{n=1} r_k (-1)^k e^{k\zeta_{cc}}, \\
    &\bar V = - \pi \sum\limits_{k=1} v_k \left(\sqrt{y^2 -1} -y\right)^k = - \pi \sum\limits_{k=1} v_k (-1)^k e^{k\zeta_{cc}}.
\end{align*}
We may now form the functions $u = V\bar R + \bar V R$ and $b = V \bar V$ and find the boundary values of analytic functions $U$ and $B$ as follows:
\begin{align*}
  &U(\eta) = \frac{1}{2\pi i} \int\limits_{-\pi}^{\pi} \frac{u(-\eta') - u(\eta')}{2} 
  \cot \frac{\eta'-\eta}{2} \, d\eta', \\
  &B(\eta) = \frac{1}{2\pi i} \int\limits_{-\pi}^{\pi} \frac{b(-\eta') - b(\eta')}{2} 
  \cot \frac{\eta'-\eta}{2} \, d\eta',
\end{align*}
which can be efficiently computed by means of the fast Fourier transform.

\subsection{Equations in $\chi$-plane and integrals of motion}

By making additional transformation~\eqref{chi1} the equations of motion become particularly 
simple, and are given by the following relations:
\begin{align}\label{EqCHI}
	& -l\sin\eta R_t - \left(\dot c + \dot l \cos\eta\right) R_{\eta} = i\left(U R_{\eta} - U_{\eta} R\right), \\
	& -l\sin\eta V_t - \left(\dot c + \dot l \cos\eta\right) V_{\eta} = i\left(U V_{\eta} - B_{\eta} R\right). 
\end{align}
 These equations have two integrals of motion which are
 equivalent to the ones found in~\cite{Tanveer1993}. Indeed, one can integrate equations (\ref{EqCHI}) using expansion (\ref{fourier_exp}). As a result, the following expressions are valid:
 \begin{align*}  
\begin{cases}
r_1 l(t) = -2Q = const \\
v_1 l(t) \,= -2\Gamma \,=const 
\end{cases}\Rightarrow \quad
e^{-i\chi} = -\frac{l(t)}{2w} \hspace{10pt} \mbox{as} \hspace{10pt} w\to\infty
\end{align*}

or
\begin{align}
&R(w) = 1 + \frac{Q}{w} +\mathcal{O}\Big(\frac{1}{w^2}\Big), 
\hspace{20pt}
V(w) = \frac{\Gamma}{w} +\mathcal{O}\Big(\frac{1}{w^2}\Big)\hspace{10pt}|w|\to\infty \label{farfield} 
\end{align}
$Q$ and $\Gamma$ are called ``winding'' and virtual ``circulation''. Conservation of circulation can 
be viewed as Kelvin's theorem for virtual fluid, and winding is related to the topology of the conformal 
plane.

\subsection{Formation of singularity in infinite time}

We shall consider the lowest order expansion so that the initial datum satisfies the constraints~\eqref{nine} and still results in a nontrivial fluid flow. The functions $R$ and $V$ at time $t=0$ are given by:
\begin{align}
    V(\xi) &= l f_1(\xi) = i\left[w - ic - \sqrt{(w-ic)^2 + l^2}\right]  \label{Vini}\\
    R(\xi) &= 1 - f_2(\xi) = \left[ 1 + \frac{1}{l^2}\left( w-ic - \sqrt{(w-ic)^2 + l^2}\right)^2 \right] \label{Rini}
\end{align}
It is assumed that at least for some time, these functions are given by convergent series of the form~\eqref{cheb_exp}, or equivalently the Fourier series~\eqref{fourier_exp}.  This is a crucial assumption 
that is only based on the results of numerical simulations of 
the short branch cut~\citep{Dyachenko2021short}, and preceding theoretical work (see~\citep{Tanveer1993,dyachenko2019poles}). Vaguely speaking this assumption will hold when the only zeros 
of $R$ in the first Riemann sheet are located at the branch points. Naturally one must ensure that the initial 
data~\eqref{Vini}-\eqref{Rini} has no additional zeros in the first sheet.

In order to establish the motion of the branch points we seek
the complex conjugated functions given by:
\begin{align}
  \tilde V &= -i\left[w+ic - \sqrt{(w+ic)^2 + l^2}\right], \\
  \tilde R &= \left[1 + \frac{1}{l^2}\left(w+ic - \sqrt{(w+ic)^2 + l^2}\right)^2 \right].
\end{align}
Then we must calculate all four functions 
$V$, $\tilde V$, $R$ and $\tilde R$ on the imaginary axis by replacing $w\to iv$. We introduce:
\begin{align}
    S &= \sqrt{(v-a)(v-b)} = \sqrt{l^2 - (v-c)^2}, \\
    F &= \sqrt{(v+a)(v+b)} = \sqrt{(v+c)^2 - l^2}
\end{align}
Now 
\begin{align}
    V &= -v + c - iS, \\
    \tilde V &= v + c - F
\end{align}
and we may write
\begin{align}
    R &= 1 - \frac{\left(v-c + iS\right)^2}{l^2} = 
    \frac{2}{l^2} \left[S^2 - i(v-c)S \right], \\
    \tilde R &=  1 - \frac{\left(v+c - F\right)^2 }{l^2} = 
    \frac{2}{l^2}F\left(v+c-F\right)
\end{align}
Then we introduce 
\begin{align}
    Q = R\tilde V + \tilde R V
\end{align}
to end up with transport velocity $U$:
\begin{align}
    U = P^- Q
\end{align}
Notice that the singularities of $Q$ in the upper half--plane are the ones coming from the function $S$. 
We can replace $Q\to Q_{ess}$ where $Q_{ess}$ only includes the terms proportional to $S$. A simple 
calculation reveals the following expression:
\begin{align}
    Q_{ess} = \frac{2i}{l^2} \left[2lFS - \left(c^2 + ab + 2cv\right)S\right]
\end{align}
Then 
\begin{align}
    U = \left(1 - i\hat H\right)Q_{ess},
\end{align}
where $\hat H$ is the Hilbert transform. 
To calculate the ``Hilbert'' transform one should remember that after projecting to the lower half--plane $Q_{ess}$ must be replaced by its analytic continuation to the upper half--plane. This is done by restoring the 
regular part by the corresponding singular part as in formula~\eqref{analytic_Cheb}. It amounts to writing:
\begin{align}
    iS  &\to iS - (v-c), \\
    ivS &\to ivS - v(v-c) + \frac{1}{2}l^2
\end{align}

Collecting all terms together we find following expression for transport velocity $U$:
\begin{align}
    U = Q_{ess} + \frac{2}{l^2}\left[ 2cv^2 + abv - c \left(c^2 + \frac{l^2}{2} + ab\right) + 2cI\right]
\end{align}
where 
\begin{align}
    I(v) = \frac{1}{\pi} v. p. \int\limits_{a}^{b} \frac{\sqrt{(s^2-a^2)(b^2-s^2)}\,ds}{s-v}
\end{align}
Now we calculate $U(a)$. As long as $S(a) = 0$ this expression is purely real. After tedious calculations we find that 
\begin{align}
    U(\varepsilon) = \frac{8b}{(1-\varepsilon)^2} \left[ 
    -\frac{1}{4} - \varepsilon + \frac{1}{2} \varepsilon^2 + \varepsilon^3 + I(\varepsilon)\right],
\end{align}
here $\varepsilon = \frac{a}{b}$ and 
\begin{align}
I(\varepsilon) = \frac{1}{\pi} \int\limits_{\varepsilon}^1\sqrt{\frac{(q+\varepsilon)(1-q^2)}{q-\varepsilon}}\,dq
\end{align}
This integral is expressed in terms of elliptic functions, but we will not provide explicit formulas. We study asymptotic behaviour of $I(\varepsilon)$ at $\varepsilon \to 0$. To do this we use the expansion:
\begin{align}
    \sqrt{\frac{q+\varepsilon}{q - \varepsilon}} = 1 + \frac{\varepsilon}{q} + \sum\limits_{k=0}^{\infty}c_k \left(\frac{\varepsilon}{q}\right)^{k+2}
\end{align}
Thus we must estimate the integral:
\begin{align}
    I(\varepsilon) = \frac{1}{\pi} \int\limits_{\varepsilon}^1 \left(1 + \frac{\varepsilon}{q} + \sum\limits_{k=0}^{\infty}c_k \left(\frac{\varepsilon}{q}\right)^{k+2}\right)\sqrt{1-q^2}\,dq.
\end{align}
Now we mention that
\begin{align}
    \frac{1}{\pi} \int\limits_0^1\sqrt{1-q^2}\,dq = \frac{1}{4}
\end{align}
One can show that 
\begin{align}
    U(\varepsilon) = 8b\left(-\frac{1}{4} + \frac{1}{4} + \frac{1}{\pi}\varepsilon\ln\varepsilon + O(\varepsilon)\right),
\end{align}
and therefore:
\begin{align}
U(\varepsilon) = \frac{8b}{\pi}\varepsilon\ln\varepsilon + O(\varepsilon),
\end{align}
where $O(\varepsilon)$ denote terms of order $\varepsilon$ and all higher orders. It is not necessary to determine them precisely.

We ended up with the following result:
\begin{align}
    U(a) \to \frac{8}{\pi}a\ln\frac{a}{b} + O(a)\,\mbox{at}\,a\to 0.
\end{align}
The differential equation:
\begin{align}
    \dot{a} = \frac{8}{\pi} a\ln\frac{a}{b}
\end{align}
with initial data:
\begin{align}
    \ln\frac{a}{b} = -\ln\frac{b}{a_0} = -c
\end{align}
gives solution 
\begin{align}
    \ln\frac{a}{b} = - c e^{\frac{8t}{\pi}},
\end{align}
and 
\begin{align}
    a = be^{-ce^{\frac{8t}{\pi}}}\,\quad c = -\ln\frac{b}{a_0}
\end{align}
It means that singularity $a=0$ is never achieved in a finite time, while it is approaching to the real line faster than exponential estimates predict. 

It is important to emphasize that this conclusion will hold only provided that no singularity from the second sheet crosses the branch cut and invalidates the assumptions of single 
pair of branch points in the first Riemann sheet. Nevertheless, 
numerical evidence suggests that at least one such solution does exist, see the Fig. $2a$ in~\citep{Dyachenko2021short}



\section{Conclusion}

A potential fluid flow in 2D Euler equations with a free boundary is considered with $V$ and $R$  
having a pair of square--root branch points $w = ia$ and $w = ib$ in conformal plane. We 
obtained the following main results:
\begin{itemize}
    \item{The equations of motion have been transplanted to the cut connecting $ia$ to $ib$ in $w$--plane,
    and introduced a new conformal mapping $\chi(w)$ that allows the study of singularities in higher Riemann sheets. With the second sheet of $w$-plane unfolded, one can study the singularities in multiple sheets, and possibly obtain new integrals of motion from pairs of cuts in higher sheets. This conjecture will have implications for a further study of integrability of water waves.}
    \item{We developed an asymptotic theory based on the exact equations formulated at the cut that 
    shows double--exponential rate of approach of singularity at $w=ia(t)$ to the fluid domain. The 
    presented theory answers a long--standing question about formation of singularity at the free surface 
    in finite vs. infinite time when no gravity or capillarity is present. However, it remains unclear whether a solution with only a single pair of branch points in the first sheet can exist for infinite time.}
    \item{The physical fluid can be complemented with a virtual fluid, which is a 
    mathematical object whose motion is defined by the analytic continuation of $R$ and $V$ into $\mathbb{C}^+$ excluding the branch cut. The expanded domain contains a virtual fluid 
    vortex sheet, for which Kelvin's theorem for circulation holds true: a ``virtual circulation'' 
    is conserved and is one of the two new invariants discovered in this work. The second invariant, 
    the ``winding'', is related to the genus of conformal plane of the virtual fluid -- the number of holes.}
\end{itemize}
The present work offers a new look at the problem of water waves and illustrates a deeply rooted 
connection between water waves and vortex sheet problem that was overlooked by the water waves 
community. Despite the fact that the present work is devoted to only a pair of square--root branch points, 
one may consider a general system of many pairs of branch points connected by branch cuts, $\gamma_k(t)$.
The equations for analytic functions $R$ and $V$ are then formed at each $\gamma_k(t)$ and fully determine 
the flow of ``virtual fluid''. Each of the branch cuts, $\gamma_k(t)$ is equivalent to a ``virtual'' 
vortex sheet, whose dynamics is governed by 2D Euler equations. An open question one could address is 
related to the rolling of a vortex sheet: Is breaking of a steep water wave related to instabilities of a ``virtual'' vortex sheet? The answer to this question is presently a 
work in progress.

Each pair of square--root branch points is associated with a new pair of integrals of motion: ``winding''
and virtual ``circulation''. A particle analogy springs into mind. Nevertheless these invariants alone 
do not fully describe the fluid flow, this follows from equations~\eqref{EqCHI}, which requires all 
Fourier modes to be defined, however 
``winding'' and ''circulation'' only define the first Fourier modes of $R$ and $V$. Nevertheless, the 
conformal domain $\chi(w)$ opens the second Riemann sheet for $R$ and $V$ which must contain 
more singularities and fully define the analytic functions $R$ and $V$ at the cuts. We conjecture that 
tracking of all singularities and their associated invariants uniquely describes the fluid flow, and 
are analogous to the ``action'' variables in the long standing problem of integrability of 2D potential
Euler equations.

\section{Acknowledgements}
Studies presented in it he subsections 2.4 and 2.5 were supported by the Russian Science Foundation 
grant no. 19-72-30028. 
The work of V.~E. Zakharov was supported by the National Science Foundation, grant no. DMS-1715323. 
The work of S.~A. Dyachenko was supported by the National Science Foundation, grants no. DMS-1716822 and no. DMS-2039071. 
This material is based upon the work supported by the National Science Foundation under Grant No. DMS-1439786 while S.~A. Dyachenko was in residence at the Institute for Computational and Experimental Research in Mathematics in Providence, RI, during a part of the Hamiltonian Methods in Dispersive and Wave Evolution Equations program.

{\bf Declaration of Interests}.  The authors report no conflict of interest.



\appendix

\section{Derivation of $U$ and $B$}\label{AppendixA:U_and_B}
We will provide a derivation for the function $B$, and a calculation for $U$ can be done 
analogously. Let us consider the product:
\begin{align}
    \bar V V = \frac{1}{\pi^2} \iint \frac{v(\tau)\bar v(\tau')d\tau d\tau'}{(\tau+iw)(\tau' - iw)}
\end{align}
and use the relation:
\begin{align}
    \frac{1}{(\tau + iw)(\tau' - iw)} = 
    \frac{1}{\tau + \tau'}\left[\frac{1}{\tau + iw} + \frac{1}{\tau' - iw}\right].
\end{align}
The product $\bar V V$ then becomes:
\begin{align}
    \bar V V = \frac{1}{\pi}\int\left[\frac{1}{\pi} \int \frac{\bar v(\tau') d\tau'}{\tau' + \tau}\right] \frac{v(\tau)d\tau}{\tau + iw} + \frac{1}{\pi}\int\left[\frac{1}{\pi} \int \frac{v(\tau) d\tau}{\tau + \tau'}\right] \frac{\bar v(\tau)d\tau'}{\tau' - iw}
\end{align}
One can observe that the terms enclosed in square bracket are given by the following:
\begin{align}
    V(-i\tau') = - \frac{1}{\pi}\int\frac{v(\tau)d\tau}{\tau + \tau'}\quad\mbox{and}\quad\bar V(i\tau) = -\frac{1}{\pi}\int\frac{\bar v(\tau')d\tau'}{\tau'+\tau}
\end{align}
and result in the following formula for $\bar V V$:
\begin{align}
    \bar V V = -\frac{1}{\pi}\int \frac{v(\tau)\bar V(i\tau)d\tau}{\tau + iw} - 
    \frac{1}{\pi}\int\frac{\bar v(\tau) V(-i\tau)}{\tau - iw} = \hat P^-\left[\bar V V\right] + \hat P^+\left[\bar V V\right],
\end{align}
Thus the function $B(w)$ is defined in the complex plane via:
\begin{align}
    B(w) = -\frac{1}{\pi} \int\frac{v(\tau)\bar V(i\tau)d\tau}{\tau + iw} = -\frac{1}{\pi}\int\frac{b(\tau)d\tau}{\tau + iw} \label{AppendixA:B}
\end{align}
where we defined $b(\tau) := v(\tau)\bar V(i\tau)$ at the interval $\tau\in[a,b]$ to be the 
jump at the branch cut. Analogous calculation for $U$ results in:
\begin{align}
    U(w) = -\frac{1}{\pi} \int \frac{u(\tau)d\tau}{\tau + iw} = 
           -\frac{1}{\pi}\int \frac{\left[v(\tau)\bar R(i\tau) + r(\tau)\bar V(i\tau) \right] d\tau}{\tau+iw}, \label{AppendixA:U}
\end{align}
where we defined $u(\tau) := v(\tau)\bar R(i\tau) + r(\tau)\bar V(i\tau)$ at the interval $\tau \in [a,b]$.

\newpage

\section{Supplemental Relations with Chebyshev Polynomials}\label{A1}

We consider an auxiliary contour integral around interval $[-1,1]$ and apply residue theorem:
\begin{align}
	I = \oint\limits_{[-1,1]} \frac{\left(T_n(x) + \sqrt{x^2 - 1}\, U_{n-1}(x)\right)\,dx}{x + y} = 
	    \oint\limits_{[-1,1]} \frac{\left(x + \sqrt{x^2-1}\,\right)^n\,dx}{x + y} ,
\end{align}
where we used the following identity for Chebyshev polynomials:
\begin{align}
	T_n(x) + \sqrt{x^2-1}\,U_{n-1}(x) = \left(x+\sqrt{x^2-1}\right)^n \label{ChebExp}.
\end{align}
We consider branch of square--root $\sqrt{x^2 - 1} = i\sqrt{1-x^2}$ above the cut, and $\sqrt{x^2 - 1} = -i\sqrt{1-x^2}$ below. 
The axiliary integral $I$ is related to integral of interest by the following formula:
\begin{align} 
	\int\limits_{-1}^{1} \frac{\sqrt{1-x^2}\,U_{n-1}(x)\,dx}{x + y} = \frac{1}{2i} \oint\limits_{[-1,1]} \frac{\left(T_n(x) + \sqrt{x^2 - 1}\, U_{n-1}(x)\right)\,dx}{\xi + y} =  \frac{I}{2i},
\end{align}
where we have noted that integral of regular function $\frac{T_n(x)}{x+y}$ around a closed curve vanishes.
The contour integral $I$ consists of four parts, however circular integrals around $x = \pm 1$ vanish because the integrand has 
no poles at these points, therefore leaving only the two integrals above and below cut:
\begin{align}
	I = \int\limits_{-1}^{1} \frac{\left(x+i\sqrt{1-x^2}\right)^n - \left(x-i\sqrt{1-x^2}\right)^n}{x+y}\,dx.
\end{align}
We make a substitution $x = \cos t$, $0<t<\pi$ and symmetrize the interval of integration:
\begin{align}
	I = \int \limits_{\pi}^{0} \frac{e^{int}-e^{-int}}{\cos(t) + y} (-\sin t) \,dt = \frac{1}{2} \int\limits_{-\pi}^{\pi}\frac{e^{int}-e^{-int}}{\cos t + y}\sin t\,dt,
\end{align}
and after the substitution $z = e^{it}$, $|z| = 1$ and $dt = \frac{dz}{iz}$:
\begin{align}
	I = -\frac{1}{2}\oint\limits_{|z|=1} \frac{\left(z^n -z^{-n}\right)\left(z-z^{-1}\right)}{z^2 + 2yz+1}\,dz.
\end{align}
The integrand has pole of order $(n+3)$ at $z = 0$ and a pair of simple poles at $z = -y \pm\sqrt{y^2-1}$. When $y = \xi + 2 + \frac{2a}{l} > 1$ and 
only the poles at $z = 0$ and $z = -y + \sqrt{y^2-1}$ lie within the unit circle. Note that if $y < -1$ the poles switch. 

\underline{{\bf Note} that poles switch if $ y < -1$! Also it is straightforward to obtain result for complex $y$. }

The residues at $z = -y \pm \sqrt{y^2-1}$ can be computed explicitly and by virtue of~\eqref{ChebExp} are given by:
\begin{align}
	\res_{-y\pm\sqrt{y^2-1}}\frac{(z^n - z^{-n})(z-z^{-1})}{z^2 + 2yz+1} = \pm2\sqrt{y^2-1}\,U_{n-1}(-y), \label{Res1}
\end{align}
where we used symmetry for Chebyshev polynomials of even and odd $n$. 

For the residue at zero we first note the generating function of Chebyshev polynomials:
\begin{align}
	\frac{(z^n - z^{-n})(z-z^{-1})}{1+2yz+z^2} = \left(z^n - \frac{1}{z^n}\right)\left(z-\frac{1}{z}\right)\sum\limits_{k=0}^\infty U_k(-y)z^k,
\end{align}
to write Laurent series, and note that only the following terms contribute to the principal part:
\begin{align}
	\frac{1}{z^{n+1}}\sum\limits_{k=0}^\infty U_k(-y)z^k - \frac{1}{z^{n-1}}\sum\limits_{k=0}^\infty U_k(-y)z^k,
\end{align}
and the residues are:
\begin{align}
\begin{cases}
	U_1(-y), &n = 1 \\
	U_n(-y) - U_{n-2}(-y), &n > 1.
\end{cases}
\end{align}
It is convenient to use recursion relations for Chebyshev polynomials $2T_n(x) = U_n(x) - U_{n-2}(x)$ and $2T_1(x) = U_1(x)$ to have the residue written
in the compact form:
\begin{align}
	\res_{z=0} \frac{(z^n - z^{-n})(z-z^{-1})}{z^2 + 2yz+1} = 2T_n(-y). \label{Res0}
\end{align}
The final result for $y>1$ is obtained by summing the two residues~\eqref{Res1} and~\eqref{Res0}:
\begin{align}
	\frac{1}{\pi}\int\limits_{-1}^{1} \frac{\sqrt{1-x^2}\,U_{n-1}(x)\,dx}{x + y} = -\left[T_n(-y) + \sqrt{y^2-1}\,U_{n-1}(-y)\right] =
		-\left(\sqrt{y^2-1} - y\right)^n, \label{CintRe}
\end{align}
\underline{{\bf Note}} that this can be extended for $-1<y<1$, and as follows:
\begin{align}
	\frac{1}{\pi} p.v. \int\limits_{-1}^{1} \frac{\sqrt{1-x^2}\,U_{n-1}(x)\,dx}{x - y} = -T_n(y).\label{CintPV}
\end{align}

\bibliography{waterwaves}

\end{document}